\documentclass[12pt]{iopart}
\usepackage{iopams} 
\usepackage{theorem}

\def \cA {{\cal A}} 
\def \cB {{\cal B}}
\def \cD {{\cal D}}
\def \cH {{\cal H}}

\def \cM {{\cal M}}
\def \cX {{\cal X}}
\def \hT {{\hat T}}

\def \C {{\mathbb C}}
\def \N {{\mathbb N}}
\def \R {{\mathbb R}}
\def\tp{\otimes}
\def\defeq{{:=}}

\def\openone{\hbox{\upshape \small1\kern-3.3pt\normalsize1}}
\def\id{{\rm id}}
\def\tr{{\rm \,tr\,}}

\newcommand \map[3]{#1 : #2 \rightarrow #3}
\newcommand \set[1] {\{ #1 \}}

\newcommand \abs[1] {| #1 |}
\newcommand \norm[1] {\| #1 \|}
\newcommand \lnorm[2] {\| #1 \|_{#2}}
\newcommand \trnorm[1] {\lnorm{#1}{1}}
\newcommand \ave[1] {\langle #1 \rangle}

\newcommand \braket[2] {\langle #1 | #2 \rangle}
\newcommand \ketbra[2] {|#1 \rangle \langle #2|}

\theorembodyfont{\upshape}
\newtheorem{theorem}{Theorem}
\newenvironment{proof}{\noindent {\bf Proof. }}{\hfill
  $\blacksquare$\\}
\newenvironment{rem}{\noindent {\bf Remark. }}{\hfill $\square$\\}

\begin{document}

\letter{Entropy production rates of bistochastic
  strictly contractive quantum channels on a matrix algebra}

\author{Maxim Raginsky\dag\footnote[3]{E-mail address:  {\tt
      maxim@ece.northwestern.edu}}}

\address{\dag\ Center for Photonic Communication and Computing,
  Department of Electrical and Computer Engineering, Northwestern
  University, Evanston, IL 60208-3118, USA}

\begin{abstract}
We derive, for a bistochastic strictly contractive quantum channel
on a matrix algebra, a relation between the contraction rate and the
rate of entropy production.  We also sketch some applications of
our result to the statistical physics of irreversible processes and to
quantum information processing.
\end{abstract}

\pacs{03.65.Yz, 03.67.-a, 05.30.-d}



\nosections

Let $\cA$ be the algebra of observables (say, a C*-algebra with
identity), associated with a quantum-mechanical system $\Sigma$.  A general
evolution of $\Sigma$ is described, in the Heisenberg picture, by a
map $\map{T}{\cA}{\cA}$ which is (i) unital: $T(\openone) =
\openone$, and (ii) completely positive:  for any nonnegative integer
$n$, the map $\map{T \tp \id}{\cA \tp \cM_n}{\cA \tp \cM_n}$, where $\cM_n$
  is the algebra of $n \times n$ complex matrices, sends positive
  operators to positive operators \cite{kra}.
We shall henceforth refer to such maps as (quantum) channels
\cite{key}.  If $\Sigma$ is an $N$-level system, then its algebra of
observables is isomorphic to $\cM_N$, which is exclusively the case we
shall consider in this letter.  A celebrated result of
Kraus \cite{kra} then says that, for any channel $T$,
there exists a collecton of at most $N^2$ operators $V_i \in \cM_N$,
which we shall call the Kraus operators associated with $T$, such that
(i) $T(A) = \sum_i V^*_i A V_i$, and (ii) $\sum_i V^*_i V_i =
\openone$.  It is now easy to see that $T(A^*) = T(A)^*$ for all $A
\in \cM_N$, i.e., any channel maps Hermitian operators to Hermitian operators.

Given a channel $T$, the corresponding
Schr\"odinger-picture channel $\hT$ is defined via the duality
$$
\tr{[\hT(A)B]} = \tr{[AT(B)]},
$$
whence it follows that $\hT$ is a completely positive map which
preserves the trace, i.e., $\tr \hT(A) = \tr A$ for all $A \in
\cM_N$.  In other words, $\hT$ maps the set $\cD_N$ of $N \times N$
density matrices into itself.  Furthermore, in terms of the Kraus operators
$V_i$, we have $\hT(A) = \sum_i
V_i A V^*_i$, so that $\hT(A^*)=\hT(A)^*$ as well.

The set $\set{T^n}_{n \in \N}$ is a discrete-time quantum dynamical
semigroup generated by $T$, i.e., $T^n T^m = T^{n+m}$ and we take $T^0
\equiv \id$ (the identity channel).  It is easy to show that, for any
channel $T$, there exists at least one density operator $\rho$ such
that $\hT(\rho)=\rho$ \cite{tdv}.  A question of clear physical
importance is to determine whether the dynamics generated by $T$ is
relaxing \cite{tdv}, i.e., whether there exists a density operator
$\rho$ such that, for any density operator $\sigma$, the orbit
$\set{\hT^n(\sigma)}$ converges to $\rho$ in the trace norm
$\trnorm{A} \defeq \tr{(A^* A)^{1/2}}$.

One way to show that a dynamics is relaxing relies on the so-called
Liapunov's direct method \cite{str}. Let $\cX$ be a
compact separable space, and let $\map{\varphi}{\cX}{\cX}$ be a
continuous map, such that
\begin{itemize}
\item[(i)] $\varphi$ has a unique fixed point $x_0 \in \cX$, and
\item[(ii)] there exists a Liapunov function for $\varphi$, i.e., a
  continuous functional $S$ on $\cX$ such that, for all $x \in \cX$,
  $S[\varphi(x)] \ge S(x)$, where equality holds if and only if $x
  \equiv x_0$.
\end{itemize}
Then, for any $x \in \cX$, the sequence $\set{\varphi^n(x)}$ converges
to $x_0$. 

Let $T$ be a bistochastic channel, i.e., one for which $T(\openone) =
\hT(\openone) = \openone$.  If we treat $\cM_N$ as a Hilbert space
with the Hilbert-Schmidt inner product, $\ave{A,B} \defeq \tr (A^*
B)$, then an easy calculation shows that the Schr\"odinger-picture
channel $\hT$ is precisely the adjoint of $T$ with respect to
$\ave{\cdot,\cdot}$, i.e., $\ave{A,T(B)} = \ave{\hT(A),B}$ for all
$A,B \in \cM_n$.  The composite map $T \circ \hT$ (which we shall
write henceforth as $T\hT$) is also a
bistochastic channel, which is, furthermore, a Hermitian operator with
respect to $\ave{\cdot,\cdot}$.  In \cite{str2}, Streater proved the
following result.

\begin{theorem}\label{th:streater}
Let $\map{T}{\cM_N}{\cM_N}$ be a bistochastic channel.  Suppose that
$\hT$ is ergodic with a spectral gap $\gamma \in [0,1)$, i.e., (i) up to
  a scalar multiple, the identity matrix $\openone$ is the only fixed
  point of $\hT$ in all of $\cM_N$, and (ii) the spectrum of $T\hT$ is
  contained in the set $[0,1-\gamma] \cup \set{1}$.  Then, for any
$\sigma \in \cD_N$, we have
\begin{equation}
S[\hT(\sigma)]-S(\sigma) \ge \frac{\gamma}{2} \lnorm{\sigma -
  N^{-1}\openone}{2}^2,
\label{eq:streater}
\end{equation}
where $S(\sigma)\defeq -\tr{(\sigma \ln{\sigma})}$ is the von Neumann entropy
 of $\sigma$ and $\lnorm{A}{2} \defeq [\tr(A^* A)]^{1/2}$ is the
Hilbert-Schmidt norm of $A$.
\end{theorem}

In other words, if a bistochastic channel $\hT$ is ergodic, then the
  dynamics generated by $T$ is relaxing by Liapunov's
  theorem\footnote[1]{Endowing the set $\cD_N$ with the trace-norm
  topology takes care of all the continuity requirements imposed by
  Liapunov's theorem.}. Furthermore, the relaxation process is
  accompanied by entropy production at a rate controlled by the spectral gap.

Now we have an interesting ``inverse'' problem. Consider a
bistochastic channel $T$ on $\cM_N$ with $\hT$ strictly contractive
\cite{rag}.  That is, $\hT$ is uniformly continuous on $\cD_N$ (in the
trace-norm topology) with Lipschitz constant $C \in [0,1)$:
  for any pair $\sigma,\sigma' \in \cD_N$, we have
  $\trnorm{\hT(\sigma) - \hT(\sigma')} \le C\trnorm{\sigma -
    \sigma'}$.  Then by the contraction mapping principle
  \cite{rs}, $N^{-1}\openone$ is the only density matrix
  left invariant by $\hT$, and furthermore $\trnorm{\hT(\sigma) -
    N^{-1}\openone} \rightarrow 0$ as $n \rightarrow \infty$ for any
  $\sigma \in \cD_N$, i.e., the dynamics generated by $T$ is relaxing.
  The question is, does the entropy-gain estimate (\ref{eq:streater})
  hold, and, if so, how does the spectral gap $\gamma$ depend on the
  contraction rate $C$?

This problem was motivated in the first place by the following
observation. In the case of $\cM_2$, the action of a bistochastic strictly
contractive channel $\hT$ can be given a direct geometric
interpretation. Recall that the density matrices in $\cM_2$ are in a
one-to-one correspondence with the points of the closed unit ball in $\R^3$.
Then the image of $\cD_2$ under a strictly contractive
channel $\hT$ with contraction rate $C$ will be contained inside the
closed ball of radius $C$, centered at the origin \cite{rag}, i.e.,
the image of $\cD_2$ under $\hT$ will consist only of mixed states.  This
geometric illustration suggests that the rate of entropy increase
under $\hT$ must be related to the contraction rate.  Now even
though in the case of $\cM_N$ with $N \ge 3$ we no longer have such
a convenient geometric illustration, nevertheless it seems plausible
that the rate of entropy production under a bistochastic strictly
contractive channel would still be controlled by the contraction rate.

Indeed it turns out that the contraction rate is related to the rate
of entropy production, as stated in the following theorem.

\begin{theorem}\label{th:myentgain}
Let $T$ be a bistochastic channel on $\cM_N$, such that $\hT$ is
strictly contractive with rate $C$.  Then $\hT$ is ergodic with
spectral gap $\gamma \ge 1 - C$, so that, for any $\sigma \in \cD_N$, we have
\begin{equation}
S[\hT(\sigma)]-S(\sigma) \ge \frac{1-C}{2}\lnorm{\sigma -
  N^{-1}\openone}{2}^2.
\label{eq:myentgain}
\end{equation}
\end{theorem}

\begin{proof} We first prove that $\hT$ is ergodic.  As we noted
  before, $T$ and $\hT$ are adjoints of each other with respect to the
  Hilbert-Schmidt inner product.  Using the Kadison-Schwarz
  inequality \cite{kad}
$$
\Phi(A^* A) \ge \Phi(A)^* \Phi(A)
$$
for any channel $\Phi$ on a C*-algebra $\cA$, as well as the fact
that
$$
\tr T(A) = \tr[\hT(\openone)A] = \tr A
$$
for any $A \in \cM_N$, we find that
$$
\lnorm{T(A)}{2}^2 = \tr[T(A)^* T(A)] \le \tr[T(A^* A)] =
\tr(A^* A) = \lnorm{A}{2}^2,
$$
and the same goes for $\hT$.  That is, both $T$ and $\hT$ are
contractions on $\cM_N$ (in the Hilbert-Schmidt norm), hence their
fixed-point sets coincide \cite{nf}.  

By hypothesis, $\hT$ leaves invariant the density matrix
$N^{-1}\openone$, which is invertible.  In this case a theorem of
Fannes, Nachtergaele, and Werner \cite{fnw,bjkw} says that $T(X)=X$ if
and only if $V_i X = XV_i$ for all $V_i$, where $V_i$ are the Kraus
operators associated with $T$.  It was shown in \cite{rag} that if
$\hT$ is strictly contractive, then the set of all $X$ such that
$V_i X = XV_i$ for all $V_i$ consists precisely of
multiples of the identity matrix.  We see, therefore, that $T(X)=X$ if
and only if $X = \chi \openone$ for some $\chi \in \C$, whence it
follows that $\hT(X) = X$ if and only if $X$ is a multiple of
$\openone$.  This proves ergodicity of $\hT$.

Our next task is to establish the spectral gap estimate $\gamma \ge 1
- C$.  Let $X$ be a Hermitian operator with $\tr{X}=0$.
In that case we can find a density operator $\rho$ and a sufficiently small
number $\epsilon > 0$ such that $\sigma \defeq \rho + \epsilon X$ is
still a density operator \cite{tdv}\footnote[2]{This may be seen as
  a simple consequence of the following fact \cite{an}. The set
  $\cD^{\rm inv}_N$ of all invertible $N \times N$ density matrices
  is a smooth manifold, where the tangent space at any $\rho \in \cD^{\rm
    inv}_N$ can be naturally identified with  the set of $N \times N$
  traceless Hermitian matrices.}. Then
\begin{equation}
\trnorm{\hT(X)} = (1/\epsilon)\trnorm{\hT(\sigma)-\hT(\rho)} \le
  (C/\epsilon)\trnorm{\sigma-\rho} = C\trnorm{X}.
\label{eq:contr}
\end{equation}
Because one is a simple eigenvalue of both $T$ and $\hT$, it is
also a simple eigenvalue of $T\hT$.  Hence $1-\gamma$ (which we may
as well assume to belong to the spectrum of $T\hT$) is the largest
eigenvalue of the restriction of $T\hT$ to traceless matrices.  Let
$Y$ be the corresponding eigenvector. Without loss of generality we
may choose $Y$ to be Hermitian\footnote[3]{Recall that $1-\gamma$ is real, and
$T\hT(A^*)=[T\hT(A)]^*$ for all $A$, which implies that $Y+Y^*$ is
also an eigenvector of $T\hT$ with the same eigenvalue.}.  Then, using
Eq.~(\ref{eq:contr}) and the fact that $\trnorm{\Phi(A)} \le
\trnorm{A}$ for any trace-preserving completely positive map $\Phi$
\cite{rus}, we may write
$$
(1-\gamma)\trnorm{Y} = \trnorm{T\hT(Y)} \le \trnorm{\hT(Y)} \le
C\trnorm{Y},
$$
which yields the desired spectral gap estimate.  The entropy gain bound
(\ref{eq:myentgain}) now follows from Theorem \ref{th:streater}.
\end{proof}

\begin{rem}Eq.~(\ref{eq:contr}) can also be proved using the following
  finite-dimensional specialization of a general result due to Ruskai
  \cite{rus}.  If $\map{T}{\cM_N}{\cM_N}$ is a channel, then
\begin{equation}
\sup_{A=A^*; \tr{A}=0} \frac{\trnorm{\hT(A)}}{\trnorm{A}} =
\frac{1}{2}\sup_{\psi,\phi \in \C^N; \braket{\psi}{\phi}=0}
\trnorm{\hT(\ketbra{\psi}{\psi}-\ketbra{\phi}{\phi})}.
\label{eq:ruskai}
\end{equation}
Because $\hT$ is strictly contractive, the right-hand side of
(\ref{eq:ruskai}) is bounded from above by $C$, and (\ref{eq:contr})
follows. The supremum on the left-hand side of (\ref{eq:ruskai}) is
the ``Dobrushin contraction coefficient,'' studied extensively by
Lesniewski and Ruskai \cite{lr} in connection with the contraction of monotone
Riemannian metrics on quantum state spaces under (duals of) quantum
channels.\end{rem}

Note that in some cases the sharper estimate
\begin{equation}
S[\hT(\sigma)] - S(\sigma) \ge
\frac{1-C^2}{2}\lnorm{\sigma-N^{-1}\openone}{2}^2
\label{eq:sharp}
\end{equation}
may be shown to hold.  Consider, for instance, the case $T=\hT$, so
that the eigenvalues of $\hT$ are all real.  Let
$\lambda_1,\ldots,\lambda_L$, $L=N^2-1$, be the eigenvalues of $\hT$
that are distinct from unity.  Then we claim that $\max_j
\abs{\lambda_j} \le C$, which can be proved via {\it reductio ad
  absurdum}.  Suppose that there exists some $X$ (which we may take to
be Hermitian) with $\tr{X}=0$ such that $\hT(X)=\lambda X$ with
$\abs{\lambda} > C$. We may use the same trick as in the proof above
to show that there exist two density operators, $\sigma$ and $\rho$,
such that $\trnorm{\hT(\sigma) - \hT(\rho)} > C\trnorm{\sigma-\rho}$,
which would contradict the strict contractivity of $\hT$.
Because $T\hT = \hT^2$, we have $1-\gamma = \left(\max_j
\abs{\lambda_j}\right)^2 \le C^2$, which confirms (\ref{eq:sharp}).
Furthermore, using a theorem of King and Ruskai \cite{kr}, the bound
(\ref{eq:sharp}) can be established for {\it all} bistochastic
strictly contractive channels on $\cM_2$, as well as for tensor
products of such channels.

The proof of this last assertion goes as follows.  Let $T$ be a
channel on $\cM_2$ such that $\hT$ is strictly contractive.  Then
$\hT$ is ergodic, the proof of which can be taken {\it verbatim} from the
proof of Theorem \ref{th:myentgain}.  It is left to show that
$1-\gamma \le C^2$. Any density operator in $\cM_2$ can be written as
\begin{equation}
\rho = \frac{1}{2}\left(\openone + \sum^3_{j=1}r_j\sigma_j\right),
\label{eq:2den}
\end{equation}
where the $\sigma_j$ are the Pauli matrices
$$
\sigma_1 = \left(\begin{array}{cc}
0 & 1 \\
1 & 0 \end{array}\right),\qquad
\sigma_2 = \left(\begin{array}{cc}
0 & -i \\
i & 0 \end{array}\right),\qquad
\sigma_3 = \left(\begin{array}{cc}
1 & 0 \\
0 & -1 \end{array}\right),
$$
and the real numbers $r_j$ satisfy the condition $r^2_1 + r^2_2 +
r^2_3 \le 1$ (this is precisely the one-to-one correspondence between
$\cD_2$ and the closed unit ball in $\R^3$).  The King-Ruskai theorem
\cite{kr} asserts that, for any bistochastic channel $T$ on $\cM_2$,
there exist unitaries $U,V$ and real numbers $\xi_j$, $1 \le j \le 3$,
with $\abs{\xi_j} \le 1$ such that, for any $\rho \in \cD_2$,
$$
\hT(\rho) = U[\hT_{\rm diag}(V\rho V^*)]U^*,
$$
where the action of the map $\hT_{\rm diag}$ on the density operator
(\ref{eq:2den}) is given by
$$
\hT_{\rm diag}(\rho) = \frac{1}{2}\left(\openone +
\sum^3_{j=1}\xi_jr_j\sigma_j\right).
$$
It can be easily shown \cite{rag} that if $\hT$ is strictly
contractive, then $C = \max_j \abs{\xi_j}$.

The parameters $\xi_j$ are determined as follows \cite{kr}.  Consider
the orthonormal basis of $\cM_2$, generated by $\openone$ and the
Pauli matrices, with respect to which $\hT$ and $T$ can be written as
$4\times 4$ matrices in the block-diagonal form
$$
\hT = \left(\begin{array}{cc}
1 & 0 \\
0 & M \end{array}\right),\qquad
T = \left(\begin{array}{cc}
1 & 0 \\
0 & M^t \end{array}\right),
$$
where $M^t$ denotes the transpose of $M$\footnote[5]{The $3\times
3$ matrix $M$ must be real because both $\hT$ and $T$ map
Hermitian operators to Hermitian operators.}.  Then the absolute values
of the parameters $\xi_j$ are precisely the singular values of $M$.
This implies that $C = \norm{M}$, where $\norm{M}$ denotes the
operator norm (the largest singular value) of $M$.  Consequently we
get $1-\gamma = \norm{M^* M} = \norm{M}^2 \equiv C^2$, which yields
the entropy gain estimate (\ref{eq:sharp}).  The argument for tensor
products of bistochastic channels on $\cM_2$ runs along similar lines.

We remark that the results reported in this letter are consistent with
the following theorem \cite{tdv}.  Let $T$ be a channel on $\cM_N$ with
the property that $\hT$ has a unique fixed point $\rho \in \cD_N$ in
all of $\cM_N$. Let $\lambda_j$ be the eigenvalues of $\hT$ distinct
from one, and let $\kappa \defeq \max_j \abs{\lambda_j}$.  Then there
exist a polynomial $p$ and an $N$-dependent constant $K$ such that,
for any $\sigma \in \cD_N$,
\begin{equation}
\trnorm{\hT^n(\sigma)-\rho} \le Kp(n)\kappa^n.
\label{eq:tdv}
\end{equation}
This shows that the dynamics generated by $T$ is relaxing,
$\trnorm{\hT^n(\sigma)-\rho} \rightarrow 0$ as $n\rightarrow \infty$,
and that the rate of convergence is controlled essentially by the
eigenvalue of $\hT$ with the second largest modulus.  Now, if $T$ is a
bistochastic channel with $\hT$ strictly contractive, then it follows
from Theorem \ref{th:myentgain} that $\kappa \le C^{1/2}$.  But by
virtue of strict contractivity we have
$$
\trnorm{\hT^n(\sigma) - N^{-1}\openone} \equiv \trnorm{\hT^n(\sigma) -
  \hT^n(N^{-1}\openone)} \le C^n\trnorm{\sigma-N^{-1}\openone} <
\frac{2C^{n/2}(N-1)}{N},
$$
which has the form of (\ref{eq:tdv}).  To obtain the last inequality we
used the fact $C < C^{1/2}$ for $0 \le C < 1$, as well as the fact
that the set $\cD_N$ is compact and convex, so that the convex
functional $\trnorm{\sigma - N^{-1}\openone}$ attains its supremum on
an extreme point of $\cD_N$, i.e., on a pure state.  In turn, a routine
calculation shows \cite{rag} that, for any pure state $\sigma$,
$\trnorm{\sigma-N^{-1}\openone} = 2(N-1)/N$.

Finally, a few comments are in order as to how our results come to
bear upon (a) the statistical physics of irreversible processes
\cite{str} and (b) quantum information processing \cite{ksv}.

In the first setting
we are interested in a concise mathematical description of the approach to
equilibrium.  We
consider a quantum-mechanical system with the Hilbert space $\cH$,
$\dim \cH \le \infty$. We will use $\cA$ to denote the algebra $\cB(\cH)$ of all
bounded linear operators on $\cH$.  Let $H$ be the (possibly unbounded)
Hamiltonian of the system, with the additional requirement that the
operator $e^{-\beta H}$ is trace-class for all positive real $\beta$.
This means that the Gibbs state exists for all positive inverse
temperatures, and that each eigenvalue of $H$ has finite
multiplicity.  Let us write down the spectral decomposition $H =
\sum_E P_E$, where $E$ are the eigenvalues of $H$ and $P_E$ are the
corresponding eigenprojections.  Then $n_E \defeq \dim P_E \cH <
\infty$ for all $E$.  Suppose we are given a channel $T : \cA
\rightarrow \cA$ such that, for each $E$, the restriction $T_E$ of $T$ to
the algebra $\cA_E \defeq P_E \cA P_E$ is bistochastic, i.e., $T(P_E)
= \hT(P_E) = P_E$, so that $T(\cA_E) \subseteq \cA_E$, and strictly
contractive.  Then Theorem \ref{th:myentgain} says that, for each $E$,
Eq.~(\ref{eq:myentgain}) holds with some contraction rate $C_E$.  The
Hilbert spaces $P_E \cH$ can be thought of as the energy levels of
$H$, and the states $P_E/n_E$ as the corresponding microcanonical
states.  If in addition we can show that $C \defeq \sup_E C_E < 1$,
then any normal state of the entire system will converge toward the
Gibbs state with the rate $C$.

In the framework of quantum information processing we are concerned mostly with
systems that have finite-dimensional Hilbert spaces.  We may envision
either a quantum register that is used for storing information, or a
quantum computer that processes information.  In both of these
situations we are interested in stabilizing the information against
the effects of decoherence. The decoherence mechanism is modelled by a
channel. It was shown in \cite{rag} that the set
of strictly contractive channels on an algebra $\cA$ is dense in the
set of all channels on $\cA$ (here we are talking about the
Schr\"odinger-picture channels).  The same result also holds for
bistochastic strictly contractive channels, which are dense in the set
of all bistochastic channels \cite{rag}.  For such channels Theorem
\ref{th:myentgain} may be used to estimate the rate at which entropy
is produced in the register or in the computer. Such estimates
generally serve as a measure of efficiency of error-correction
procedures \cite{ncsb} (note that it was shown in \cite{rag} that errors modelled
by strictly contractive channels can be corrected only
approximately).

\ack This research was supported
in part by the U.S. Army Research Office through the MURI (Multiple
Universities Research Initiative) program, and in part by the Defense
Advanced Research Projects Agency through the QuIST (Quantum
Information Science and Technology) program.

\end{document}